\documentclass{PoS}

\title{Nuclear modification of full jets and jet structure in relativistic nuclear collisions}

\ShortTitle{Nuclear modification of full jets and jet structure}

\author{\speaker{Ning-Bo Chang}$^{a,b}$, Guang-You~Qin$^{b}$, Yasuki~Tachibana$^{c}$   \\
        \llap{$^{a}$}Institute of Theoretical Physics, Xinyang Normal University, Xinyang, Henan 464000, China\\
        \llap{$^{b}$}Institute of Particle Physics, Central China Normal University, Wuhan 430079, China\\
        \llap{$^{c}$}Department of Physics and Astronomy, Wayne State University, Detroit, Michigan 48201, USA\\
        E-mail: \email{changnb@xynu.edu.cn, guangyou.qin@mail.ccnu.edu.cn, yasuki.tachibana@wayne.edu}}
        
\abstract{With our coupled jet-fluid model, we study the nuclear modifications of full jets and jet structures for single inclusive jet and $\gamma$-jet events in relativistic heavy-ion collisions. The evolution of full jet showers is studied via a set of coupled transport equations including the effects of collisional energy loss, transverse momentum broadening and medium-induced splitting. The dynamics of the jet energy and momentum deposited into the medium is described by hydrodynamic equations with source terms. Our detailed analysis indicates that collisional absorption (energy loss) tends to narrow the jet shape function while transverse momentum kicks and medium-induced radiations broaden the jet transverse profile. Also, jet-induced flow plays a significant contribution to jet shape function and dominates at large angles away from the jet axis. The final nuclear modification pattern for the jet shape function is a combined effect from various jet-medium interaction mechanisms. Our detailed studies for single inclusive jets and $\gamma$-jets for various kinematics indicate that the nuclear modification of jet shape has strong dependence on jet energy and collision energy, and weak dependence on jet flavor (quark or gluon).}

\FullConference{International Conference on Hard and Electromagnetic Probes of High-Energy Nuclear Collisions\\
		30 September - 5 October 2018\\
		Aix-Les-Bains, Savoie, France}

\begin{document}

\section{Introduction} 
  Jet quenching is a powerful tool to study the property of the quark gluon plasma (QGP) created in relativistic heavy-ion collisions. When jets propagate through the QGP medium, the collisions between jet partons with medium partons can reduce the energy of jet partons and therefore modify the energy and internal structure of full jets. Earlier studies of jet quenching mainly focus on the quenching of high $p_T$ hadrons produced from the leading parton of full jets. The studies on the suppression of the spectra of high $p_T$ hadrons indicate that leading partons lose energy mainly via inelastic collisions, i.e., medium induced radiations \cite{Qin:2007rn}.
  
 Since full jets can be reconstructed using data from heavy ion collision experiments, more and more studies shift to the reconstruction of full jets because their modification can provide more observables and hence provide more detailed information about jet medium interactions.  One important difference between the quenching of full jets and leading partons is that the mechanism of full jets energy loss may be different from leading partons. On one hand, radiated partons from leading parton can be still in the jet cone which reduces the energy loss from medium induced radiation. On the other hand, while leading partons lose little energy via elastic collisions, there are many partons in a jet and each of them may suffer elastic collisions with the medium which enhances the effect of collisional energy loss. In addition the suppression of jet spectra or high $p_T$ hadron spectra which mainly probes the total energy loss thus can not discriminate between the contribution from different mechanisms. However the modification of jet internal structure may be more sensitive to different jet-medium interaction mechanisms. In our works \cite{Chang:2016gjp,Tachibana:2017syd,Chang:2019}, we study the nuclear modifications of full jet energy and jet structures for single inclusive jet and $\gamma$-jet events in PbPb collisions at both 2.76~ATeV and 5.02~ATeV, and consider the effect of medium response to jet energy and jet structure.
\vspace*{-0.3cm}

\section{Framework} \vspace*{-0.2cm}
We study full jets via the three-dimensional momentum distributions of quarks and gluons contained in the jets, $f_i(\omega_i, k_{i\perp}^2)=dN_i(\omega_i, k_{i\perp}^2)/d\omega_i dk_{i\perp}^2$, with $\omega_i$ the energy of parton $i$ and $k_{i\perp}$ its transverse momentum with respect to the jet axis. The evolution of $f_i(\omega_i, k_{i\perp}^2)$ follows the differential transport equation,
 \begin{eqnarray}
\label{eq:dG/dt2}
\hspace{-0.05in}\frac{d}{dt}f_i(\omega_i, k_{i\perp}^2, t)\hspace{-0.05in}&=&\hspace{-0.08in}\left(\hat{e}_i \frac{\partial}{\partial \omega_i}
 + \frac{1}{4} \hat{q}_i {\nabla_{k_\perp}^2}\right)\hspace{-0.05in}f_i(\omega_i, k_{i\perp}^2, t)+ \sum_j\int d\omega_jdk_{j\perp}^2 \frac{d\tilde{\Gamma}_{j\rightarrow i}(\omega_i, k_{i\perp}^2|\omega_j, k_{j\perp}^2)}{d\omega_i d^2k_{i\perp}dt} f_j(\omega_j, k_{j\perp}^2, t)\nonumber \\
&-&\sum_j \int d\omega_jdk_{j\perp}^2 \frac{d\tilde{\Gamma}_{i\rightarrow j}(\omega_j, k_{j\perp}^2|\omega_i, k_{i\perp}^2)}{d\omega_j d^2k_{j\perp}dt} f_i(\omega_i, k_{i\perp}^2, t), 
\end{eqnarray}
where the first and second terms take into account the effect of collisional energy loss and transverse momentum broadening due to elastic scatterings, while the last two terms represent the processes of medium induced radiations with $\frac{d\tilde{\Gamma}_{j\rightarrow i}}{d\omega d^2k_{\perp}dt}$ the splitting kernels taken from the higher twist formalism \cite{Wang:2001ifa}. Note that all splitting channels are included, so the evolutions of the quark distributions and the gluon distributions are coupled with each other. 

After solving for the distribution $f_i(\omega_i, k_{i\perp}^2)$ we can construct many observables, such as the energy of jets with cone size $R$, $E_{jet}(R) = \sum_i \int_R \omega_i f_i(\omega_i, k_{i\perp}^2) d\omega_i dk_{i\perp}^2$, and jet shape function $\rho_{\rm jet}(r) =\frac{1}{p_T^{\rm jet}}\frac{1}{\delta r} \sum_i p_T^i(r_i)|_{|r_i - r|\leq\frac{1}{2}\delta r}$.  In heavy-ion collisions, jet induced flow has further contributions to jet energy and jet structure. In elastic scatterings, energy is transferred from jet partons to the medium. The deposited energy will induce medium excitations, and should be included in the reconstruction of full jets. To take into account the medium response effect, we solve the hydrodynamic equation $\partial_{\mu}T_{\rm QGP}^{\mu \nu}\!\!\left(x\right)=J^{\nu}\!\!\left(x\right)$ with the following source terms,
 \begin{eqnarray}
J^{\nu}\!\!\left(x\right)&=&  -\sum_j \int \!d^3\!k_j {k^\nu_j} \left.\frac{d\!f_j \!\left(\mathbf{k}_j,\!t\right)}{d t}\!\right|_{\rm col.}
       \delta^{(3)}\left(\mathbf{x}\!-\!{\mathbf{x}}_0^{\rm jet}\!\!-\!\frac{\mathbf{k}_j}{\omega_j}t\!\right)\!\!.
\end{eqnarray}
By comparing the results with and without source terms, we obtain the jet-induced medium flow \cite{Tachibana:2017syd}. Its effect on jet energy and jet structure will be discussed in Sect.\ref{3}. 
\vspace*{-0.2cm}

\section{Numerical results}\vspace*{-0.2cm}
\label{3}
To solve the coupled differential transport equations as Eq. \ref{eq:dG/dt2}, the initial distribution should be provided. We generate it using Pythia and tune the parameters to make sure that the jet shape function in pp collisions can be described. The parameters $\hat{e}$ and $\hat{q}$ in Eq. \ref{eq:dG/dt2} are related via $\hat{q} = 4T \hat{e}$, and the value of $\hat{q}$ is obtained from the local temperature and flow velocity of the QGP medium, $\hat{q} (\tau,\vec{r}) = \hat{q}_0 \cdot \frac{T^3(\tau,\vec{}r)}{T_{0}^3(\tau_{0},\vec{0})} \cdot \frac{p\cdot u(\tau, \vec{r})}{p_0}$, with $\hat{q}_0$ the parameter to be fixed by experimental data.

\begin{figure}[tbhp]
\begin{center}
  \centering
     \includegraphics[width=0.39\linewidth]{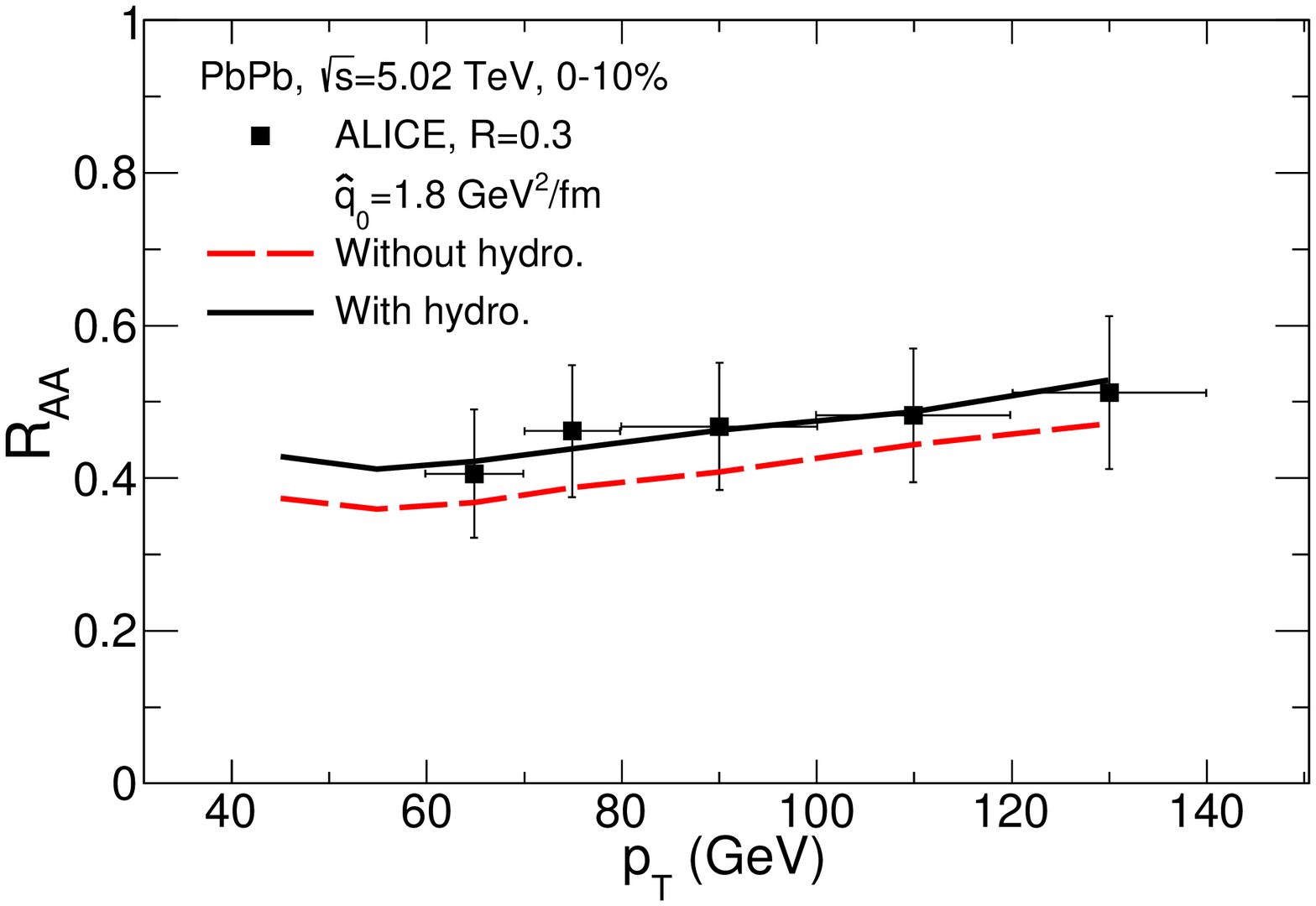}
     \includegraphics[width=0.39\linewidth]{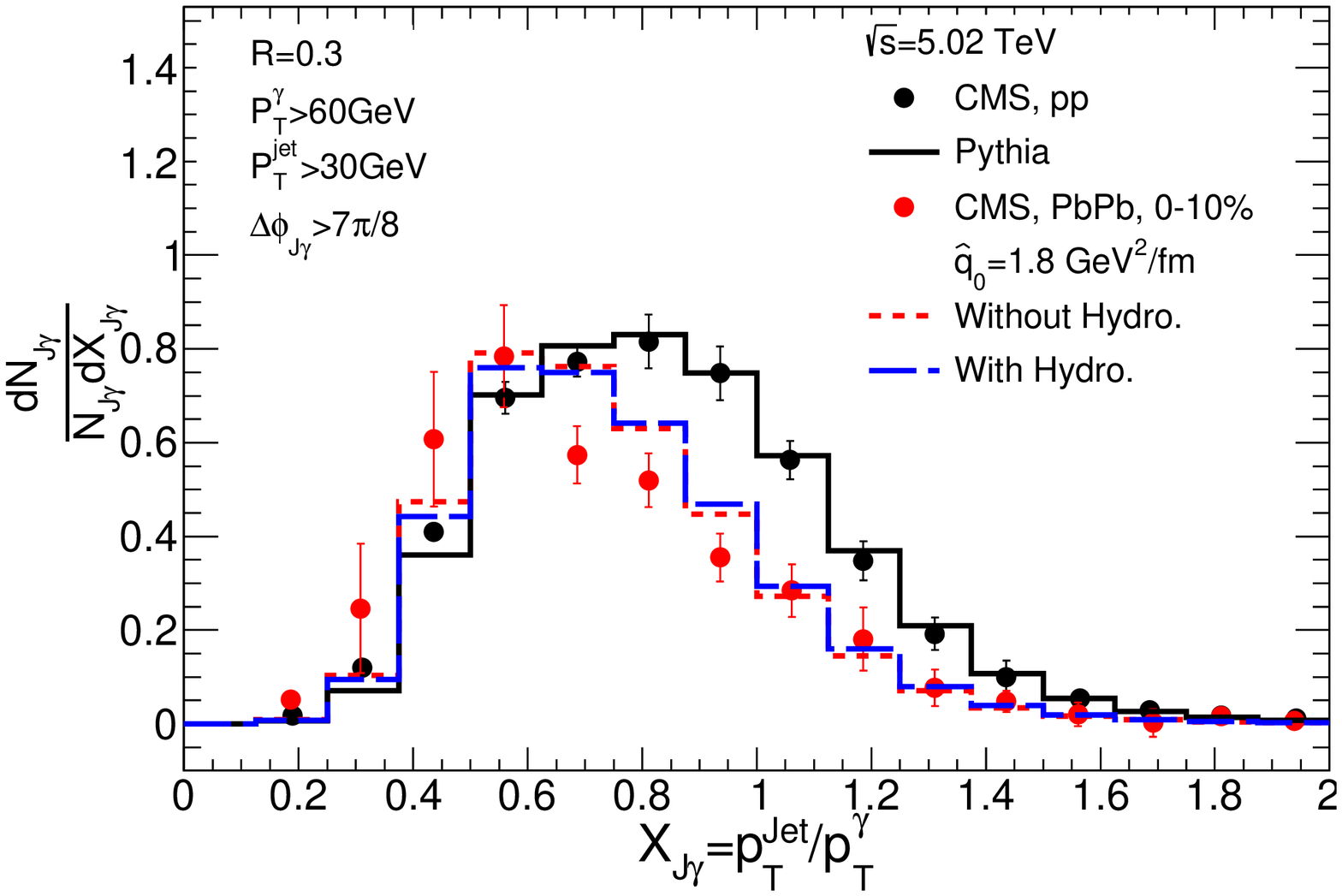}\vspace*{-0.4cm}
  \caption{Jet $R_{AA}$ (left) and the nuclear modification for $\gamma$-jet (right) momentum fraction in central Pb+Pb collisions at 5.02~ATeV. Note that `with/without hydro.' means with or without the effect of medium response. Experiment data comes from Refs. \cite{ALICE:2017,CMS:2018ogb}.}\vspace*{-0.6cm}
  \label{fig:eloss}
\end{center}
\end{figure}

After solving the transport equations, we calculate many observables in PbPb collisions and obtain their modification compared with that in pp collisions. In this proceeding, we present the results related to jet energy loss and jet structure modification in inclusive jet and $\gamma$-jet events. Fig. \ref{fig:eloss} shows the observables due to jet energy loss, including the jet $R_{AA}$ and the modification of $\gamma$-jet momentum imbalance. Both the results with or without the effect of medium response are shown. From the left panel of Fig. \ref{fig:eloss} we can see that with medium response the value of $R_{AA}$ increases, because medium response can feed back some energy into the jet cone \cite{Tachibana:2017syd}. The same conclusion can be obtained from the right panel of Fig. \ref{fig:eloss}, in which we can see that the inclusion of medium response moves the distribution of $X_{J\gamma}\equiv\frac{p_T^{jet}}{p_T^{\gamma}}$ towards larger $X_{J\gamma}$.

Our work \cite{Chang:2016gjp} analyzes the effect of each jet-medium interaction mechanism on jet energy loss and the modification of jet shape, indicating that the collisional energy loss of full jet showers contributes most to full jet energy loss while medium induced radiation contributes least. Also, our analysis shows that while medium induced radiation and transverse momentum broadening broaden the jet shape, only collisional energy makes the jet profile steeper because the hard core of the jets loses less fraction of energy than the periphery of jets. Only the combination of all three mechanisms can explain the non-monotonic behavior of $\frac{\rho(r)^{PbPb}}{\rho(r)^{pp}}$ as a function of $r$ for jets with $p_T>100$~GeV measured by CMS \cite{Chatrchyan:2013kwa} at 2.76~ATeV. In \cite{Chang:2016gjp} we predict that as jet transverse momentum decreases, one may observe a monotonic modification pattern for $\rho(r)$ as a function of $r$. The reason is that the inner core of the jet becomes softer and thus easier modified by the medium as jet energy decreases. The prediction is consistent with the new measurement by CMS in $\gamma$-jet events \cite{CMS:2018ogb} at 5.02~ATeV.
\begin{figure}[tbhp]
  \centering
     \includegraphics[width=0.8\linewidth]{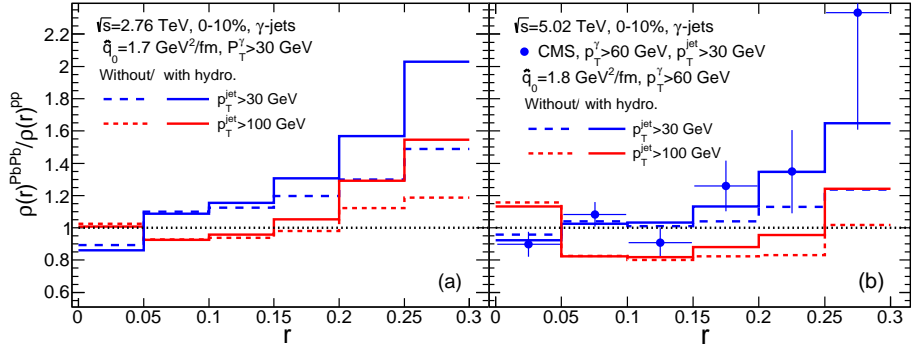}\vspace*{-0.4cm}
  \caption{Nuclear modification of the jet shape function in $\gamma$-jet events at 2.76~ATeV and 5.02~ATeV for $p_T>30$~GeV and $p_T>100$~GeV. Experiment data comes from Ref. \cite{Sirunyan:2018ncy}.}\vspace*{-0.4cm}
  \label{fig:gamma}
\end{figure}

\begin{figure}[tbhp]
  \centering
     \includegraphics[width=0.8\linewidth]{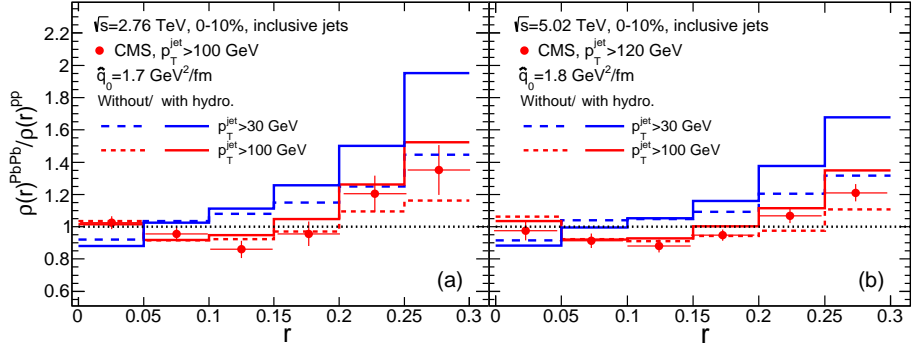}\vspace*{-0.4cm}
  \caption{The same as Fig. \ref{fig:gamma} except for inclusive jet events. Experiment data comes from Refs. \cite{Chatrchyan:2013kwa,Sirunyan:2018jqr}.}\vspace*{-0.2cm}
  \label{fig:single}
\end{figure}

  We study the nuclear modification of the jet shape function systematically using various kinematic cuts. Fig. \ref{fig:gamma} shows the modification of jet shape function in $\gamma$-jet events at 2.76~ATeV and 5.02~ATeV for jets with $p_{T}$ cut 30 GeV or 100 GeV. Fig. \ref{fig:single} shows the same results for inclusive jet events. In Fig. \ref{fig:gamma} and Fig. \ref{fig:single}, the results with and without the effect of medium response are both shown. We can see that medium response rises the value of $\frac{\rho(r)^{PbPb}}{\rho(r)^{pp}}$ at large $r$ because medium response can increase the value of $\rho(r)$ and becomes dominant at large $r$ \cite{Tachibana:2017syd}. But we can see that current experimental data can not distinguish whether the inclusion of medium response is necessary due the large errors.  We need more precise measurements to confirm the role of medium response. 
  
  Now let us look at the sensitivity of the ratio $\frac{\rho(r)^{PbPb}}{\rho(r)^{pp}}$ to jet energy, collision energy and jet flavor. From all panels in Fig. \ref{fig:gamma} and Fig. \ref{fig:single}, we can see that the results for $p_T>100$~GeV is dramatically different with $p_T>$~30 GeV. For results with $p_T$ cut 100 GeV, $\frac{\rho(r)^{PbPb}}{\rho(r)^{pp}}$ is a non-monotonic function of $r$, but for that with $p_T$ cut 30 GeV $\frac{\rho(r)^{PbPb}}{\rho(r)^{pp}}$ increases with $r$ monotonically. By comparing Fig. \ref{fig:gamma}(a) with Fig. \ref{fig:gamma}(b) and Fig. \ref{fig:single}(a) with Fig. \ref{fig:single}(b), we can see that $\frac{\rho(r)^{PbPb}}{\rho(r)^{pp}}$ is also sensitive to the collision energy \cite{Chang:2019}. Finally, by comparing Fig. \ref{fig:gamma}(a) with Fig. \ref{fig:single}(a) and Fig. \ref{fig:gamma}(b) with Fig. \ref{fig:single}(b), we can see that $\frac{\rho(r)^{PbPb}}{\rho(r)^{pp}}$ is similar in $\gamma$-jet and inclusive jet events, which means $\frac{\rho(r)^{PbPb}}{\rho(r)^{pp}}$ is not very sensitive to jet flavor \cite{Chang:2016gjp}. 
\vspace*{-0.3cm}

\section{Summary} \vspace*{-0.2cm}
We have studied full jet quenching in PbPb collisions at 2.76~ATeV and 5.02~ATeV at the LHC. We formulate a set of coupled transport differential equations for quark and gluon three dimensional momentum distributions in jets to study  the jet shower evolution in the QGP medium, in which the processes of collisional energy loss, transverse momentum broadening and medium induced radiation are included. We have also included the effect of jet-induced flow to the study of jet energy and jet structure.

 Our results can describe the experimental datas related to full jet energy loss and nuclear modification of the jet shape function in $\gamma$-jet and inclusive jet events at both 2.76~ATeV and 5.02~ATeV. The analysis indicates that the collisional energy loss dominates the full jet energy loss and plays an important role in the modification of jet shape. Our study has shown that the modification of jet shape has strong dependence on jet energy and collision energy. The modification in inclusive jet events is similar to that in $\gamma$-jet events. Jet-induced flow can feed back some energy to the jet cone and its contribution to jet shape is important at large $r$. Further precise measurements can provide more precise test to our understanding of jet quenching in heavy-ion collisions. 

\vspace{8pt}
\textbf{Acknowledgements:} 
This work is supported in part by the Natural Science Foundation of China (NSFC) under Grant Nos. 11775095, 11375072 and 11647066. N.-B. C. is supported by Nanhu Scholar Program for Young Scholars of XYNU and the CCNU-QLPL Innovation Fund (Grant No. QLPL2016P01). Y. T. is supported in part by a special award from the OVPR at Wayne State University and in part by the US NSF under Grant No. ACI-1550300.  \vspace*{-0.3cm}


\end{document}